\documentclass[5p]{elsarticle}

\usepackage{lineno,hyperref,color}
\modulolinenumbers[5]

\journal{Physica E}

\usepackage{amsmath}

\begin{document}

\begin{frontmatter}

\title{Electron waiting times for the mesoscopic capacitor}

\author{Patrick P. Hofer}
\author{David Dasenbrook}
\address{D\'epartement de Physique Th\'eorique, Universit\'e de Gen\`eve, 1211
  Gen\`eve, Switzerland}

\author{Christian Flindt}
\address{Department of Applied Physics, Aalto University, 00076 Aalto, Finland}

\begin{abstract}
We evaluate the distribution of waiting times between electrons emitted by a driven mesoscopic capacitor. Based on a wave packet approach we obtain analytic expressions for the electronic waiting time distribution and the joint distribution of subsequent waiting times. These semi-classical results are compared to a full quantum treatment based on Floquet scattering theory and good agreement is found in the appropriate parameter ranges. Our results provide an intuitive picture of the electronic emissions from the driven mesoscopic capacitor and may be tested in future experiments.
\end{abstract}


\end{frontmatter}


\section{Introduction}

The scientific career of Markus B\"uttiker was devoted to mesoscopic physics. Over several decades he made profound contributions to the field. He worked on the scattering theory of electronic conductors~\cite{buttiker85,buttiker86}, dephasing~\cite{buttiker86b,buttiker88b}, edge channel transport~\cite{buttiker88,buttiker88c}, dynamic conductors~\cite{buttiker93,moskalets02}, shot noise~\cite{buttiker90,buttiker92,blanter00} and full counting statistics~\cite{pilgram03,nagaev04}, electronic entanglement~\cite{samuelsson03,samuelsson04}, fluctuation relations~\cite{foerster08,sanchez10}, quantum heat engines~\cite{sanchez11,bergenfeldt14}, Majorana fermions~\cite{li12,jacquod13}, as well as many other topics. Early in his career he became interested in the traversal time for tunneling through a potential barrier~\cite{buttiker82,buttiker83}. He returned to questions concerning time in quantum mechanics in some of his last papers, where he investigated the waiting times between electronic transfers in mesoscopic conductors~\cite{albert11,albert12,dasenbrook14}.

During a sabbatical at University of Basel, he initiated his research on the mesoscopic capacitor \cite{buttiker93a,pretre96,gopar96,brouwer97}. The aim was to develop a quantum analogue of an $RC$-circuit. Classically, a capacitor in series with a resistor discharges on a characteristic time scale given by the $RC$-time, $\tau_{RC} = RC$. The quantum analogue proposed by Markus B\"uttiker and his co-workers consists of a small cavity connected to an electron reservoir via a quantum point contact (QPC). One might expect this capacitor to discharge over the $RC$-time given by the geometric capacitance $C$ of the cavity and the QPC resistance $R_Q/T$, where $R_Q=h/e^2$ is the resistance quantum and $T$ the transmission. Nevertheless, due to phase coherence throughout the system, the situation turns out to be different.

At low frequencies the setup still behaves like an $RC$-circuit. However, for small voltage excitations, the charge relaxation resistance
is universal and independent of the QPC transmission. In particular, for a single-channel QPC the charge relaxation resistance is simply half the value of the resistance quantum
\begin{equation}
R_\mathcal{B} = \frac{h}{2e^2}.
\end{equation}
This prediction has been verified experimentally~\cite{gabelli06}. Following a suggestion at the 27th International Conference on Low Temperature Physics, we refer to this charge relaxation resistance as the B\"uttiker resistance~\cite{glattli14talk}. The resistance quantization is robust against interactions in the cavity~\cite{mora:2010}. By contrast, the capacitance, and thus the charge relaxation time, depend on the QPC transmission.

The physics changes if a large gate voltage is applied~\cite{moskalets08}. For a square pulse excitation, the charge relaxation resistance approaches the QPC resistance and the capacitance becomes independent of the transmission~\cite{parmentier12}. In addition, the capacitor can be operated as an on-demand source of coherent
electrons. By lifting a filled level of the capacitor above the Fermi energy of the external reservoir, a single electron can be emitted from the capacitor. As the level is lowered, the capacitor is refilled with an electron. By doing so periodically, a quantized AC current can be generated~\cite{feve07,moskalets08,bocquillon13}. This has paved the way for quantum optics experiments with electrons~\cite{bocquillon14}. In particular, Hanbury-Brown-Twiss~\cite{bocquillon12} and Hong-Ou-Mandel~\cite{bocquillon13} interferometry with electrons has been realized.

In this article we characterize the mesoscopic capacitor operated as a coherent single-electron emitter by calculating the distribution of electron waiting times~\cite{albert11,albert12,dasenbrook14,brandes08,welack09,rajabi13,thomas13,albert14,thomas14,haack14,sothman14,tang14,dasenbrook15}. As such, our work combines two of Markus B\"uttiker's interests: the mesoscopic capacitor and electron waiting times. The electronic waiting time distribution (WTD) has previously been evaluated for the driven mesoscopic capacitor using a rate equation description~\cite{albert11}. Here we address the problem using two complementary methods. We use a wave packet approach to evaluate the WTD analytically together with a method based on Floquet scattering theory for numerically exact calculations. In the appropriate parameter ranges, we recover the results obtained from the rate equation description. In addition, the Floquet calculations allow us to consider parameter regimes which so far have not been accessible. The results presented here may be tested in future experiments on the mesoscopic capacitor.

The remainder of the article is organized as follows. In Sec.~\ref{sec:scat}, we discuss the mesoscopic capacitor and its description based on Floquet scattering theory. Section~\ref{sec:wtd} is devoted to the theory of electronic WTDs. In Sec.~\ref{sec:wavepacket} we use a wave packet description to analytically evaluate the WTD and the joint distribution of waiting times.  In Sec.~\ref{sec:floquet} we perform full Floquet scattering theory calculations of the WTD. Finally, in Sec.~\ref{sec:conc} we present our conclusions.

\begin{figure}
  \centering
  \includegraphics[width=0.95\columnwidth]{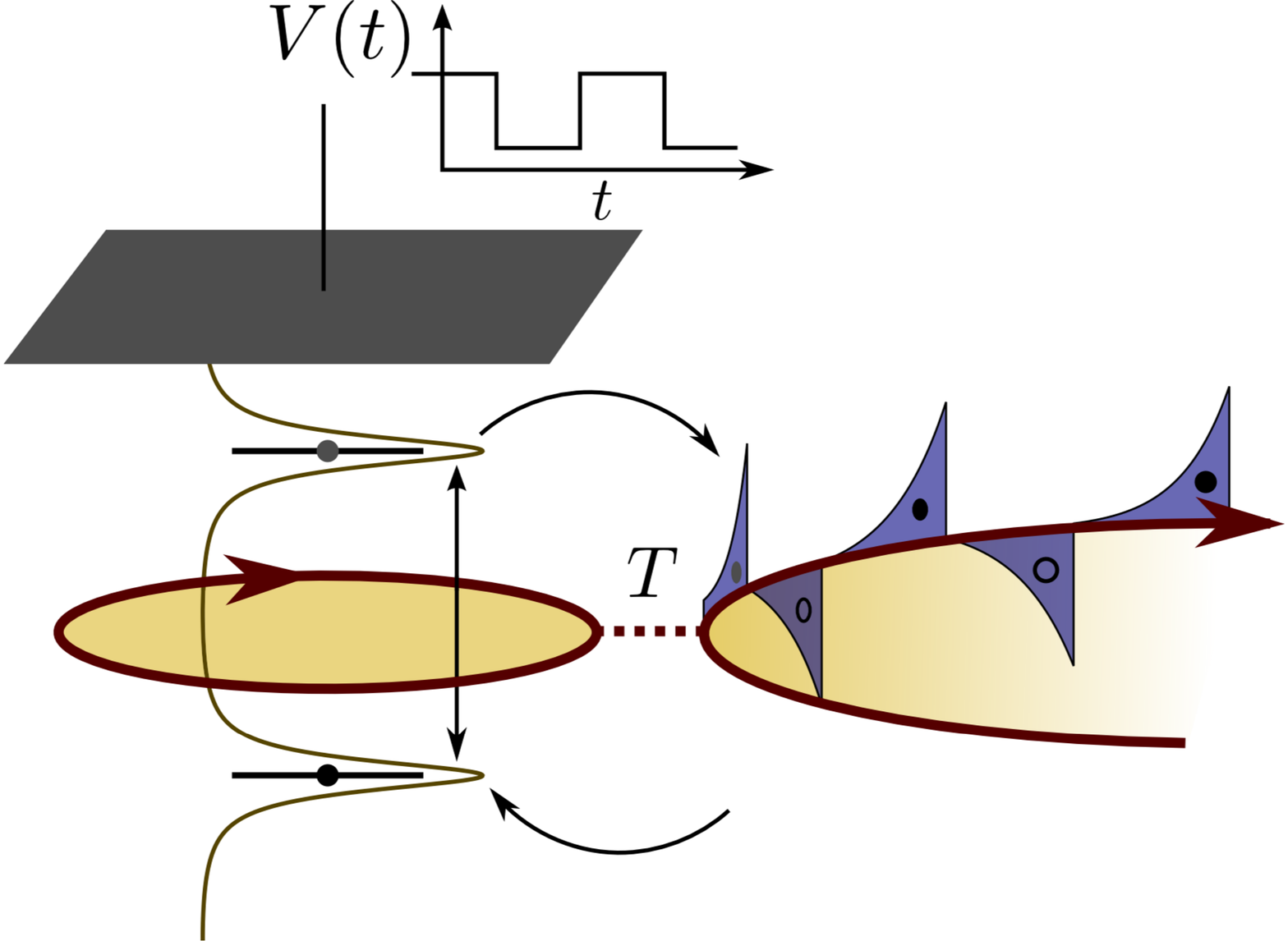}
  \caption{(color online). The driven mesoscopic capacitor. A small cavity is tunnel coupled to an edge state (in red) via a quantum point contact with transmission probability $T$. By applying a time-dependent voltage $V(t)$ to the top gate of the capacitor (in grey) the levels of the capacitor are shifted up and down. If a filled level is moved above the Fermi energy of the external reservoir, the capacitor can emit an electron into the edge state. The empty level is then refilled by an electron as the level is moved below the Fermi energy.  This leads to exponential current pulses (in blue) with alternating signs in the outgoing edge state.}
  \label{fig:mescap}
\end{figure}

\section{\label{sec:scat} The mesoscopic capacitor}
In this section we describe the mesoscopic capacitor using non-interacting scattering theory \cite{moskalets11book}. This approach can account for many experimental observations.

The capacitor is shown schematically in Fig.~\ref{fig:mescap}. Electrons propagate along chiral edge states that form when a two-dimensional electron gas is subject to a strong magnetic field. A small loop is tunnel coupled to an edge state via a QPC. The loop constitutes one plate of the capacitor. The other plate is a top-gate with potential $V(t)$. An electron traveling along the edge state can either be reflected on the capacitor and continue its motion along the edge, or it can be transmitted into the capacitor and make several turns inside the loop before eventually escaping. Since there is only a single incoming and outgoing channel, the scattering matrix of the capacitor is just a complex number of unit length which can be obtained by summing up the quantum mechanical amplitudes for all possible scattering paths.

We first consider the capacitor with a constant top-gate potential $V(t)=V_0$. The reflection
amplitude is then
\begin{align}
  \label{eq:reflectionphase}
  \mathcal{S}(\epsilon) &= r - T
\sum_{n=1}^\infty r^{n-1} e^{i n (\epsilon - e V_0) \tau_0 / \hbar} \nonumber\\
  &= \frac{r - e^{i (\epsilon - e
V_0) \tau_0 / \hbar}}{1 - r e^{i (\epsilon - e V_0) \tau_0 / \hbar}},
\end{align}
where $r$ is
the reflection amplitude of the QPC (here chosen to be real), $T = 1-r^2$, and $\tau_0 =
\ell / v_F$ is the time it takes an electron with Fermi velocity $v_F$ to complete one turn in the loop of
circumference $\ell$. In this case, no (additional) current is generated in the outgoing edge state.

To generate an AC current, the top-gate potential must be time-dependent. To see this, it is instructive to consider the density of states of the capacitor without driving
\begin{equation}
  \label{eq:dos}
  \begin{aligned}
  \rho(\epsilon) =& \frac{1}{2\pi i} \mathcal{S}^\ast(\epsilon) \frac{d
    \mathcal{S}(\epsilon)}{d \epsilon}\\
    =&\frac{1}{\pi}\sum\limits_{j=-\infty}^{\infty}\frac{\frac{1}{2}\gamma}{\left(\epsilon-eV_0-j\Delta\right)^2+\left(\frac{1}{2}\gamma\right)^2}.
    \end{aligned}
\end{equation}
The density of states consists of a series of Lorentzian peaks with spacing $\Delta = h / \tau_0$ and widths
\begin{equation}
\label{eq:widths}
\gamma=-\frac{\Delta}{\pi}\ln\left(\sqrt{1-T}\right)\simeq \frac{\Delta T}{2\pi},\hspace{.5cm}T\ll 1.
\end{equation}
A time-dependent potential $V(t)$ will shift the positions of the levels. If a filled level is moved above the Fermi energy, an electron can escape the capacitor and when an empty level is moved below the Fermi energy, an electron can be absorbed, or equivalently, a hole be emitted from the capacitor. Moving an energy level periodically above and below the Fermi energy will thus result in the periodic emission of a single electron followed by a single hole. If this is done adiabatically, the Lorentzian width of the level determines the wave-function of the emitted particles and the current consists of a series of Lorentzian pulses with alternating signs~\cite{keeling08}. Experimentally, a square-shaped potential has been applied to a capacitor with  period $\mathcal{T}$ and peak-to-peak amplitude equal to the level spacing $\Delta$~\cite{feve07}. Within one period, the potential has the form
\begin{equation}
  \label{eq:potentialstep}
  eV(t) = \begin{cases}
    \Delta/2 & 0 < t \leq \mathcal{T}/2 \\
    -\Delta/2 & \mathcal{T}/2 < t \leq \mathcal{T}
  \end{cases}.
\end{equation}
The current pulses then have the shape of decaying exponentials with alternating signs. We focus throughout this work on this square-shaped potential.

To describe a periodic potential we use Floquet scattering theory~\cite{moskalets11book}. We start by noting that an electron that completes $n$ turns in the capacitor picks up the phase~\cite{parmentier12}
\begin{equation}
  \label{eq:dynamicphase}
  \phi_n(t) = \frac{e}{\hbar} \int_{t - n \tau_0}^t V(\tau) d \tau,
\end{equation}
upon leaving the capacitor at time $t$. By substituting this expression for the static phase $(e / \hbar) V_0 n \tau_0$ in the first line of Eq.~\eqref{eq:reflectionphase}, we obtain the mixed energy-time representation of the scattering
amplitude $\mathcal{S}(t,\epsilon)$ for electrons that enter the capacitor with energy $\epsilon$ to leave it at time $t$ \cite{moskalets08}. Since the potential is periodic, $V(t)=V(t+\mathcal{T})$, we can expand the scattering phase in a Fourier series as
\begin{equation}
  \label{eq:dynamicphasefourierseries}
  e^{-i \frac{e}{\hbar} \int_0^t V(\tau) d \tau} = \sum_{m=-\infty}^{\infty} c_m e^{-i m \Omega t},
\end{equation}
where $\Omega = 2\pi / \mathcal{T}$ is the frequency of the driving. In addition, by Fourier transforming $\mathcal{S}(t,\epsilon)$ as
\begin{equation}
  \mathcal{S}_F(\epsilon_n,\epsilon) = \frac{1}{\mathcal{T}}\int_{0}^{\mathcal{T}} e^{in\Omega t}S(t,\epsilon) dt,
\end{equation}
we obtain the Floquet scattering matrix
\begin{equation}
  \label{eq:floquetsmatrix}
  \mathcal{S}_F(\epsilon_n,\epsilon) = \sum_{m=-\infty}^{\infty} c_{m+n}c^\ast_m \mathcal{S}(\epsilon_{-m}) ,
\end{equation}
where we have defined the energies
\begin{equation}
\epsilon_n = \epsilon + n \hbar \Omega.
\end{equation}
The Floquet scattering matrix contains the quantum mechanical amplitudes for an electron with energy $\epsilon_n$ to change its energy to $\epsilon$ as a result of the scattering process. Due to the periodicity of the potential, an electron can only absorb or emit energy in units of the energy quantum $\hbar \Omega$.

Equation~\eqref{eq:floquetsmatrix} is valid for an arbitrary periodic top-gate potential $V(t)$ expressed in terms of its Fourier components $c_m$. The Floquet scattering matrix relates the annihilation operators $\hat{b}(\epsilon)$ of the outgoing electrons to the annihilation
operators $\hat{a}(\epsilon)$ of the incoming electrons as
\begin{equation}
\label{eq:flocscatop}
\hat{b}(\epsilon_n) = \sum_{n=-\infty}^{\infty} \mathcal{S}_F(\epsilon_n,
\epsilon) \hat{a}(\epsilon).
\end{equation}
With this expression one may calculate the time-dependent current or the finite-frequency noise of the capacitor \cite{moskalets11book}.

The mesoscopic capacitor works best as a single-electron source if the density of states is symmetric around the Fermi energy. In this case, an analytic expression for the Floquet scattering matrix has been derived~\cite{moskalets13}. In addition, the charge relaxation resistance
\begin{equation}
\label{eq:rq}
R_q=\frac{h}{e^2}\left(\frac{1}{T}-\frac{1}{2}\right)\simeq\frac{h}{e^2T},\,\,\ T\ll 1
\end{equation}
equals the resistance of the QPC for small transmissions~\cite{parmentier12}. By contrast, the capacitance becomes independent of the transmission
\begin{equation}
\label{eq:cq}
C_q=\frac{e^2}{\Delta}.
\end{equation}
The dwell (or relaxation) time of the capacitor reads
\begin{equation}
\label{eq:dwell}
\tau_{D}=R_qC_q\simeq\frac{h}{T\Delta }=\frac{\tau_0}{T}.
\end{equation}
The capacitor is expected to operate as a nearly perfect single-electron emitter if the dwell time is (much) shorter than the period, $\tau_D\ll\mathcal{T}$. In this case, a single electron and a single hole should be emitted in almost every cycle. However, even under these optimal conditions, there can be noise at finite frequencies associated with the uncertainty of the emission time within a period and the shape of the wave packets. This type of noise has been investigated theoretically and experimentally in Refs.~\cite{mahe10,albert10,parmentier12}.

If the QPC transmission is too low, the dwell time can become comparable to the period. In this case, cycle-missing events may occur, where the capacitor fails to emit an electron within a period or an empty level is not refilled. An electron and a hole will then be missing from the otherwise periodic stream of particles. Several methods have been employed to assess the accuracy of the mesoscopic capacitor as a single-electron source. These include analyzing the finite-frequency noise \cite{moskalets08,parmentier12,moskalets13noise} or the full counting statistics of emitted charge \cite{albert10}. In a complementary approach, one considers the distribution of electron waiting times \cite{albert11}. This will be our focus in this paper.

\section{\label{sec:wtd} Electron waiting times}

The electron waiting time $\tau$ is the time that passes between two subsequent electrons being emitted by the capacitor. This is a fluctuating quantity which can be described by a probability distribution $\mathcal{W}(\tau)$. As we will see, the distribution of waiting times provides us with a useful characterization of the mesoscopic capacitor operating as a single-particle emitter. In earlier work, a rate equation description was used to evaluate the WTD \cite{albert11}. Here we address the problem using two complementary approaches. First, we calculate the WTD analytically based on
the wave packets emitted by the capacitor. These results are then compared with full Floquet scattering calculations, employing the methods developed in Refs.~\cite{dasenbrook14,dasenbrook15}. The theoretical framework for calculating WTDs using scattering theory is described in detail in Refs.~\cite{albert12,haack14,dasenbrook14,dasenbrook15}, and in this section we only provide a brief summary.

To evaluate the WTD, we relate it to the idle time probability (ITP). This is the probability that no electrons are emitted during a given time interval $[t^s,t^e]$. For time-dependent problems the ITP, $\Pi(t^s,t^e)$, depends both on $t^s$ and $t^e$. However, for a periodic process, the ITP depends only on the length of the interval $\tau=t^e-t^s$,  if the starting point $t^s$ is chosen randomly, and it can be written as
\begin{equation}
  \label{eq:itpaveraged}
  \Pi(\tau) = \frac{1}{\mathcal{T}}\int_0^\mathcal{T} \Pi(t^s,t^s +\tau) dt^s.
\end{equation}
The WTD can then be expressed as \cite{albert12,dasenbrook14}
\begin{equation}
  \label{eq:wtditp}
  \mathcal{W}(\tau) = \langle \tau \rangle \partial_\tau^2 \Pi(\tau),
\end{equation}
where $\langle \tau \rangle$ is the mean waiting time. Similar relations hold for the level spacing statistics of random matrices \cite{haake}.

The ITP can be evaluated by introducing the operator
\begin{equation}
  \label{eq:qoperator}
  \widehat{Q} = \int_{v_Ft^s}^{v_F t^e} \hat{b}^\dagger(x) \hat{b}(x) d x,
\end{equation}
where $\hat{b}(x)$ annihilates electrons at position $x$. The ITP can then be written as
\begin{equation}
  \label{eq:itp}
  \Pi(t^s,t^e) = \left \langle : e^{- \widehat{Q}} : \right \rangle,
\end{equation}
where $: \dots :$ denotes normal-ordering of operators with respect to the Fermi sea as we are only interested in electrons emitted above the Fermi level \cite{dasenbrook14}. Here we have made use of the linear dispersion relation close to the Fermi level, implying that the probability of finding no emitted electrons in the spatial interval $[v_Ft^s,v_Ft^e]$ after the capacitor is equal to the probability that the capacitor has not emitted any electrons in the temporal interval~$[t^s,t^e]$. To ease the notation, we now set $v_F=1$.

The derivatives in Eq.~(\ref{eq:wtditp}) have a physical meaning. To see this, we differentiate the ITP in Eq.~\eqref{eq:itp} with respect to the end time $t^e$ and define
\begin{equation}
  \label{eq:fptd1}
\begin{split}
  \mathcal{F}(t^s, t^e) &= -\partial_{t^e} \Pi(t^s,t^e)\\
  &= \left \langle \hat{b}^\dagger(t^e) : e^{-\widehat{Q}} : \hat{b}(t^e) \right \rangle.
\end{split}
\end{equation}
Comparing this expression with Eq.~\eqref{eq:itp}, we see that the expectation value of the normal-ordered exponential is now taken with respect to the many-body state with an electron removed at $t^e$. This is the probability distribution for an electron to be emitted at time $t^e$, given that no electrons were emitted in the interval $[t^s,t^e)$. Such a distribution is also known as a first passage time distribution. The derivative with respect to $t^e$ corresponds to the emission of an electron from the capacitor at time $t^e$. Again, for a periodic process where the starting
point is chosen randomly, we can define a first passage time distribution
\begin{equation}
  \label{eq:fptdaveraged}
  \begin{split}
  \mathcal{F}(\tau) &= \frac{1}{\mathcal{T}} \int_0^\mathcal{T}
  \mathcal{F}(t^s,t^s+\tau) d t^s\\
  &= -\partial_\tau\Pi(\tau)
  \end{split}
\end{equation}
that depends only on the interval length $\tau=t^e-t^s$.

Following a similar line of arguments, we can express the WTD as
\begin{equation}
  \label{eq:wtditp1}
  \begin{split}
  I(t^s) \mathcal{W}(t^s, t^e) =& -\partial_{t^e}\partial_{t^s}  \Pi(t^s,t^e)\\
  =& \left \langle \hat{b}^\dagger(t^e) \hat{b}^\dagger(t^s) : e^{-\widehat{Q}} : \hat{b}(t^s) \hat{b}(t^e) \right \rangle,
  \end{split}
\end{equation}
where $I(t^s)$ is the average electronic current emitted by the capacitor at time $t^s$. The expectation value yields the joint probability density for the emission of an electron both at~$t^s$ and at~$t^e$ with no electrons being emitted in between. Dividing this quantity by $I(t^s)$, we recover the conditional probability density for the first emission of an electron to occur at time $t^e$ following an emission at time $t^s$. The WTD corresponding to Eq.~\eqref{eq:wtditp} can then be expressed as
\begin{equation}
  \label{eq:wtdaveraged}
  \mathcal{W}(\tau) = \frac{\langle \tau \rangle}{\mathcal{T}} \int_0^\mathcal{T} I(t^s)
  \mathcal{W}(t^s,t^s+\tau) d t^s.
\end{equation}

Building on this principle, we can also evaluate the joint distribution of successive waiting times following our recent work described in Ref.~\cite{dasenbrook15}. To this end, we introduce a generalized ITP
\begin{equation}
\Pi(t_1^s,t_1^e;t_2^s,t_2^e)=\left<:e^{-\widehat{Q}_1-\widehat{Q}_2}:\right>,
\end{equation}
which yields the probability that the capacitor does not emit any electrons in the two time intervals~$[t_1^s,t_1^e]$~and~$[t_2^s,t_2^e]$ (which may overlap) with corresponding projection operators
\begin{equation}
  \widehat{Q}_1 = \int_{t_1^s}^{t_1^e} \hat{b}^\dagger(x) \hat{b}(x) d x,
\end{equation}
  and
\begin{equation}
  \widehat{Q}_2 = \int_{t_2^s}^{t_2^e} \hat{b}^\dagger(x) \hat{b}(x) d x.
\end{equation}
For the joint waiting time distribution, we can then write
\begin{equation}
\begin{aligned}
\label{eq:wait2t}
I(t^s)&\mathcal{W}(t^s,t^m,t^e)=\partial_{t^s}\partial_{t^e}\partial_{t_2^e}\left.\Pi(t^s,t^e;t_2^s,t_2^e)\right|_{t_2^s=t_2^e=t^m}\\
&=\left<\hat{b}^\dag(t^s)\hat{b}^\dag(t^m)\hat{b}^\dag(t^e):e^{-\widehat{Q}}:\hat{b}(t^e)\hat{b}(t^m)\hat{b}(t^s)\right>
\end{aligned}
\end{equation}
with the projector $\widehat{Q}$ defined in Eq.~(\ref{eq:qoperator}). For a periodically driven conductor, we obtain the joint distribution of subsequent waiting times just as in Eq.~\eqref{eq:wtdaveraged},
\begin{equation}
  \label{eq:wtd2taveraged}
  \mathcal{W}(\tau_1, \tau_2) = \frac{\langle \tau \rangle}{\mathcal{T}} \int_0^\mathcal{T} I(t^s)
  \mathcal{W}(t^s,t^s+\tau_1,t^s+\tau_1+\tau_2) dt^s.
  \nonumber
\end{equation}

The expressions presented in this section form the basis for our calculations of the distribution of electron waiting times using Floquet scattering theory. However, as we will now see they can also be used to evaluate the WTD for the mesoscopic capacitor using a wave packet description.

\section{Wave packet description}
\label{sec:wavepacket}
In this section we calculate the WTD for the mesoscopic capacitor based on the emitted wave packets. We consider the waiting time between emitted electrons, noting that the emitted holes in principle can also be included~\cite{albert11}. We first discuss the WTD for the capacitor when operated under ideal conditions before including unwanted effects like cycle-missing events.

\subsection{Ideal operating conditions}
\label{sec:ideal}
When operated under ideal conditions, the mesoscopic capacitor emits exactly one electron per cycle with a squared wave packet of the form
\begin{equation}
\label{eq:wavmescap}
|\psi(t)|^2=\frac{e^{-t/\tau_D}}{\tau_D}\frac{\Theta(t)\Theta\left(\frac{\mathcal{T}}{2}-t\right)}{1-e^{-\frac{\mathcal{T}}{2\tau_D}}}.
\end{equation}
The step functions reflect that the capacitor can only emit an electron in the time interval $[0,\mathcal{T}/2]$, when the energy level of the capacitor is above the Fermi level of the edge state. The dwell time $\tau_D$ is defined in Eq.~\eqref{eq:dwell} and we have ignored any possible fine structures on the time scale of $\tau_0$~\cite{moskalets08,keeling08,sasaoka10}. In addition, the normalization factor $1-e^{-\frac{\mathcal{T}}{2\tau_D}}$ ensures that the squared wave function integrates to unity. The ITP is now the product of the probabilities of not emitting any of the electrons in the time interval~$[t^s,t^e]$
\begin{equation}
\label{eq:itpid}
\Pi_{I}(t^s,t^e)=\prod\limits_{n=-\infty}^{\infty}\left(1-\int\limits_{t^s}^{t^e}|{\psi(t-n\mathcal{T})}|^2dt\right).
\end{equation}
The subscript $I$ is meant to remind us that this expression holds under ideal operating conditions.

Using the ITP we can evaluate the first passage time distribution and the WTD. Without loss of generality we set
$t^s\in[0,\mathcal{T}]$ and $t^e\ge t^s$. From Eq.~\eqref{eq:fptd1}, we then obtain
\begin{equation}
\label{eq:ttfptdid}
\mathcal{F}_{I}(t^s,t^e)=
\begin{cases}
|\psi(t^e)|^2& t^e\in[0,\mathcal{T}/2],\\
|\psi(t^e-\mathcal{T})|^2\int\limits_{0}^{t^s}|\psi(t)|^2dt & t^e\in[\mathcal{T},3\mathcal{T}/2],\\
0 &\mathrm{otherwise,}
\end{cases}
\nonumber
\end{equation}
This expression has a simple interpretation. If $t^e<\mathcal{T}$, the time $t^e$ lies within the same period as $t^s$ and the probability for the first detection to occur at $t^e$ equals the squared wave function $|\psi(t^e)|^2$ for the electron emitted in this period. No electrons can be detected in the interval $(\mathcal{T}/2,\mathcal{T})$. The first detection after $t^s$ can only happen in the second period if the electron emitted in the first period was observed before $t^s$. This gives rives to the integral in the second line above. Finally, the first passage time distribution is zero for intervals longer than $3\mathcal{T}/2$, since at least one electron is emitted in each period.

\begin{figure}
  \centering
  \includegraphics[width=\columnwidth]{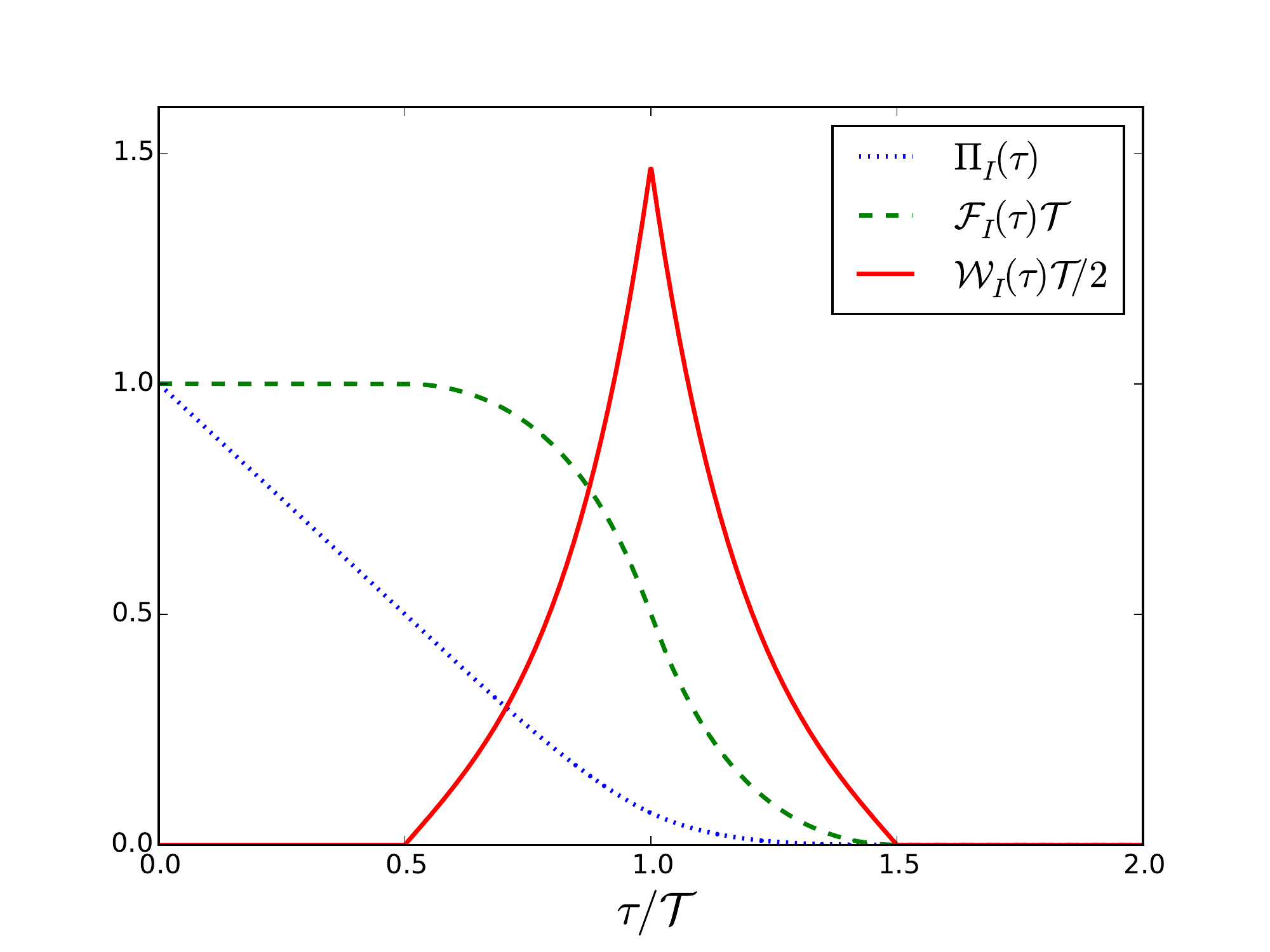}
  \caption{(color online). Analytic results for the ITP, the first passage time distribution, and the WTD for the mesoscopic capacitor. The capacitor is operated under optimal conditions, where no cycle missing events occur.  Note that the WTD is scaled down by a factor of two. The dwell time is $\tau_D=0.2\mathcal{T}$.}
  \label{fig:idealsource}
\end{figure}

For the WTD we find from  Eq.~\eqref{eq:wtditp1}
\begin{equation}
\label{eq:ttwtdid}
I_{I}(t^s)\mathcal{W}_{I}(t^s,t^e)=|\psi(t^s)|^2|\psi(t^e-\mathcal{T})|^2,
\end{equation}
with $I_{I}(t^s)=|\psi(t^s)|^2$. This is the joint probability of detecting an electron at $t^s$ followed by the next detection at $t^e$ (note that $|\psi(t^e-\mathcal{T})|^2=0$ for $t^e>3\mathcal{T}/2$). Since the wave functions are non-overlapping, the expressions for the distributions above become very simple.

Finally, we integrate over $t^s$ following Eqs.~(\ref{eq:itpaveraged}), (\ref{eq:fptdaveraged}) and (\ref{eq:wtdaveraged}). The expressions for the ITP and the first passage time distribution are somewhat cumbersome and can be found in \ref{app:stdist}. For the WTD, we obtain
\begin{equation}
\label{eq:otwtdid}
\mathcal{W}_{I}(\tau)=\frac{e^{-\frac{\mathcal{T}}{2\tau_D}}}{\tau_D}\frac{\sinh\left(\frac{\frac{\mathcal{T}}{2}-|\tau-\mathcal{T}|}{\tau_D}\right)}{\left(1-e^{-\frac{\mathcal{T}}{2\tau_D}}\right)^2}\hspace{.25cm}{\rm for }\hspace{.25cm}\tau\in\left[\frac{\mathcal{T}}{2},\frac{3\mathcal{T}}{2}\right],
\end{equation}
and zero otherwise. Here we have used that the mean waiting time is equal to the period of the drive
\begin{equation}
\label{eq:meanwtd}
\langle \tau \rangle_{I} =\mathcal{T}.
\end{equation}

In Fig.~\ref{fig:idealsource} we show the ITP, the first passage time distribution, and the WTD. The WTD gives a particularly clear picture of the emission of electrons from the capacitor. The distribution is peaked around one period of the driving with the width governed by the uncertainty in the exact time of emission within each period. The symmetry around $\tau=\mathcal{T}$ reflects that every electron is emitted with the same wave function. According to Eqs.~\eqref{eq:wtdaveraged} and~\eqref{eq:ttwtdid}, the WTD is determined by the convolution of $\left|\psi(t)\right|^2$ with $\left|\psi(t+\tau-\mathcal{T})\right|^2$. The convolution depends only on $|\tau-\mathcal{T}|$, leading to the observed symmetry.

\begin{figure*}
  \centering
  \includegraphics[width=1\columnwidth]{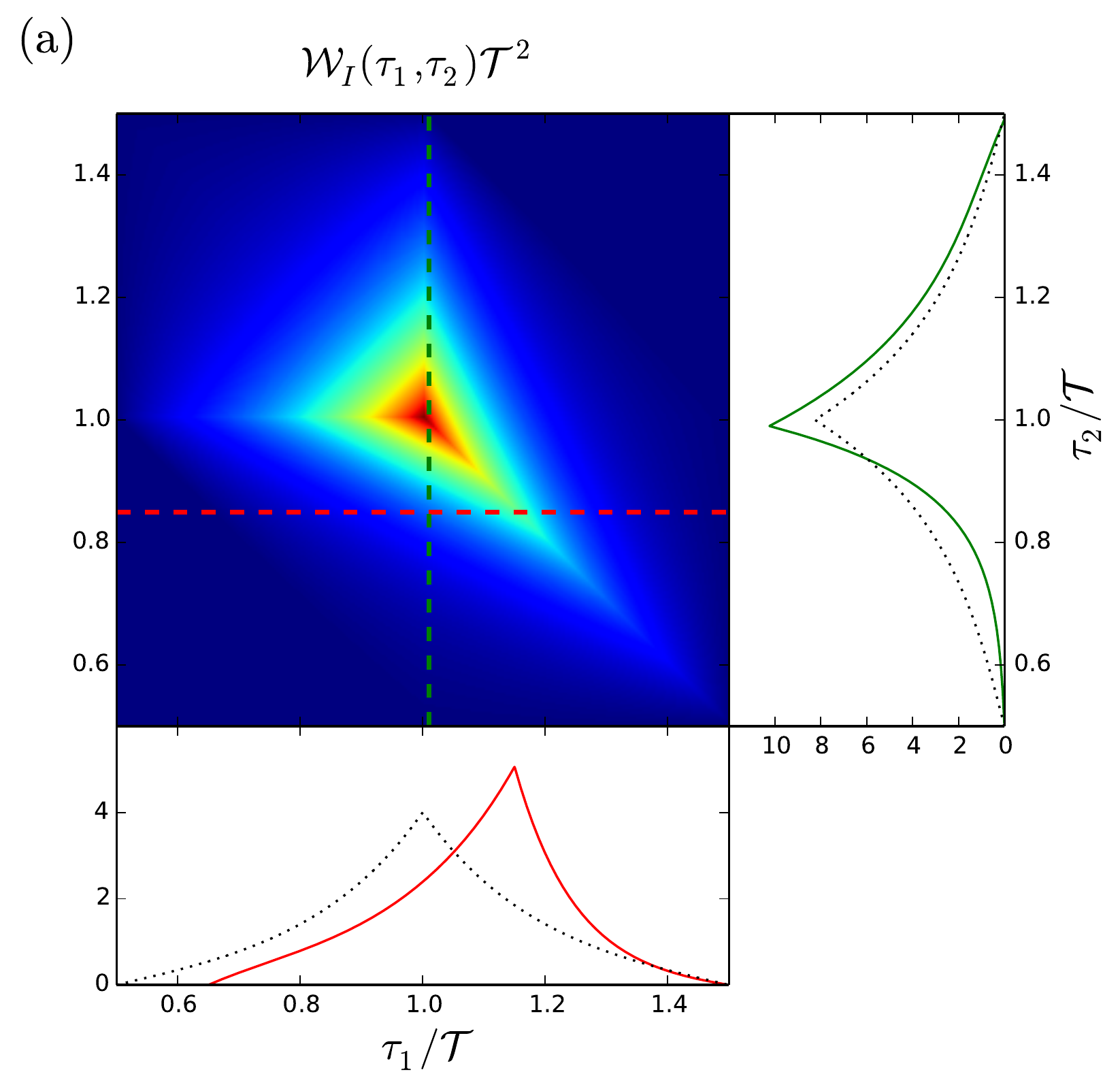}\includegraphics[width=1\columnwidth]{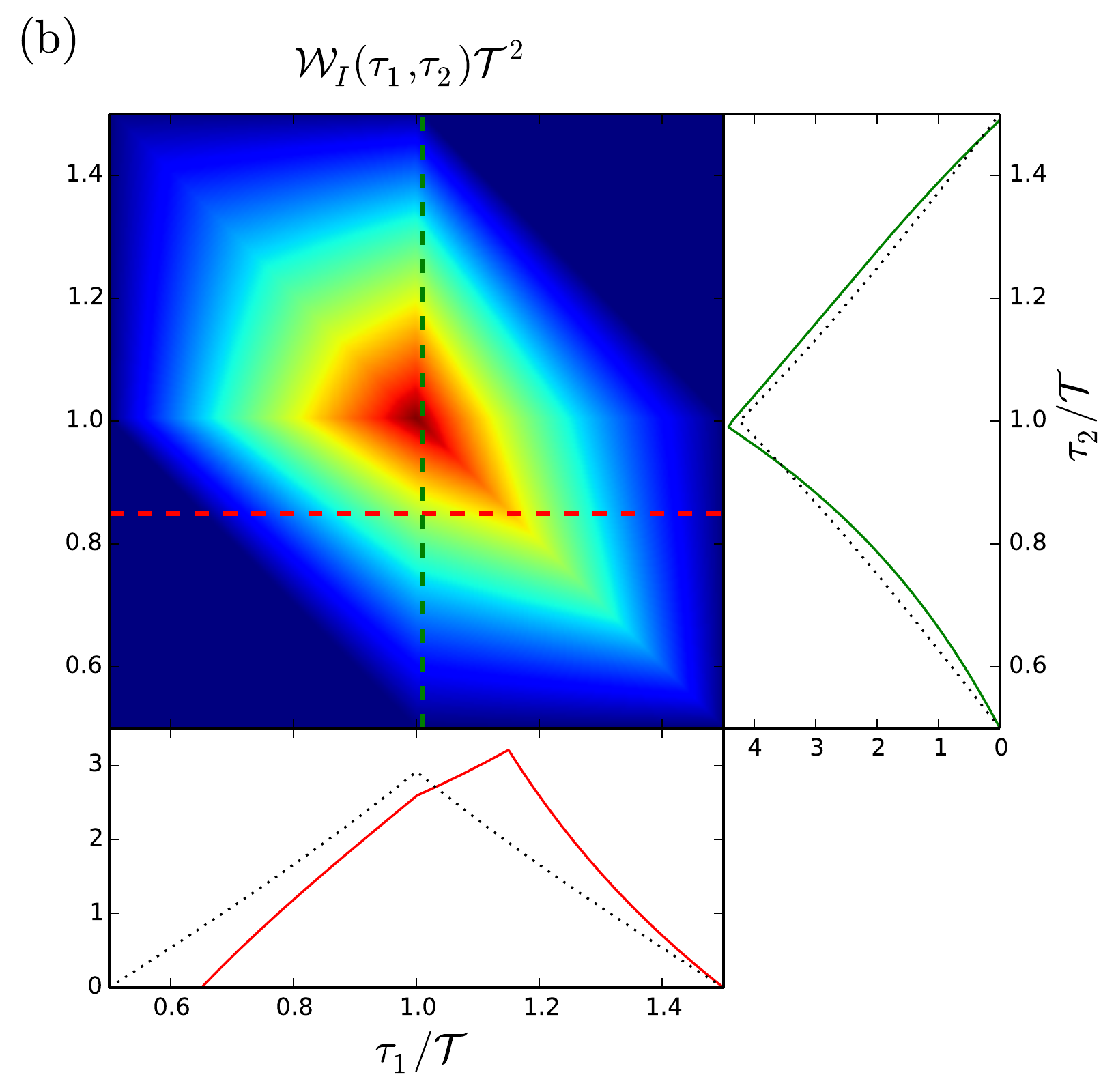}
  \caption{(color online). Analytic results for the joint distribution of waiting times. The mirror symmetry along $\tau_1+\tau_2=2\mathcal{T}$ occurs since all emitted wave packets are identical. Since the emitted wave packets have no inversion symmetry, there is no $\tau_1\leftrightarrow\tau_2$ symmetry. The side panels show the joint WTD along the cuts in the main panel. The dotted lines show the distributions if there were no correlations between subsequent waiting times. The dwell time is $\tau_D=0.2\mathcal{T}$ in (a) and $\tau_D=0.7\mathcal{T}$ in (b).}
  \label{fig:wtd2tid02}
\end{figure*}

We can also evaluate the joint WTD. The joint WTD follows from the generalized ITP which reads
\begin{align}
\label{eq:itp2tid}
\nonumber
&\Pi_{I}(t^s,t^e;t_2^s,t_2^e)=\prod\limits_{n\neq 1}^{\infty}\left(1-\int\limits_{t^s}^{t^e}|{\psi(t-n\mathcal{T})}|^2dt\right)\\&\times\bigg(1-\int\limits_{t^s}^{t^e}|{\psi(t-\mathcal{T})}|^2dt-\int\limits_{t_2^s}^{t_2^e}|{\psi(t-\mathcal{T})}|^2dt\bigg),
\end{align}
where we have chosen the interval $[t_2^s,t_2^e]$ to lie within both intervals $[t^s,t^e]$ and $[\mathcal{T},2\mathcal{T}]$.
From Eq.~\eqref{eq:wait2t}, we find the joint WTD
\begin{equation}
\label{eq:wait2tid}
I_{I}(t^s)\mathcal{W}_{I}(t^s,t^m,t^e)=|\psi(t^s)|^2|\psi(t^m-\mathcal{T})|^2|\psi(t^e-2\mathcal{T})|^2,
\nonumber
\end{equation}
which generalizes Eq.~\eqref{eq:ttwtdid} in a straightforward manner. Interestingly, this expression can be factorized as
\begin{equation}
\label{eq:wait2tidfac}
I_{I}(t^s)\mathcal{W}_{I}(t^s,t^m,t^e)=I_{I}(t^s)\mathcal{W}_{I}(t^s,t^m)\mathcal{W}_{I}(t^m,t^e),
\nonumber
\end{equation}
indicating that subsequent waiting times are uncorrelated if we can keep track of the driving of the capacitor. Again, by integrating over $t^s$, we obtain
\begin{align}
\label{eq:wtd2id}
&\mathcal{W}_{I}(\tau_1,\tau_2)= \nonumber\\
&\qquad \begin{cases}
w(\tau_1,\tau_2)&\tau_1+\tau_2\in\left[\frac{3\mathcal{T}}{2},2\mathcal{T}\right],\\
w(2\mathcal{T}-\tau_2,2\mathcal{T}-\tau_1)&\tau_1+\tau_2\in\left[2\mathcal{T},\frac{5\mathcal{T}}{2}\right],\\
0 &\mathrm{otherwise,}
\end{cases}
\end{align}
having defined
\begin{equation}
\label{eq:wtd2idhelp}
w(\tau_1,\tau_2)=\frac{e^{-\frac{3\mathcal{T}}{4\tau_D}}}{3\tau_D^2}\frac{e^{\theta_{12}/\tau_D}-e^{-\theta_{21}/\tau_D}}{\left(1-e^{-\frac{\mathcal{T}}{2\tau_D}}\right)^3},
\end{equation}
and
\begin{equation}
\label{eq:theta}
\theta_{ij}=\tau_i-\frac{\mathcal{T}}{4}-|{\mathcal{T}-\tau_j}|\left[1+\Theta(\mathcal{T}-\tau_j)\right].
\end{equation}

Figure~\ref{fig:wtd2tid02} shows the joint WTD for two different dwell times. The individual waiting times are still restricted to the interval $[\mathcal{T}/2,3\mathcal{T}/2]$. The mirror symmetry along $\tau_1+\tau_2=2\mathcal{T}$ is analogous to the symmetry around $\tau=\mathcal{T}$ for the WTD in Fig.~\ref{fig:idealsource}. In contrast to single-electron excitations created by Lorentzian voltage pulses, the joint WTD of the mesoscopic capacitor does not exhibit the symmetry $\tau_1\leftrightarrow\tau_2$. As discussed in Ref.~\cite{dasenbrook15}, this symmetry is guaranteed only if the emitted wave packets have an inversion symmetry which is not the case here. Looking at the side panels we see that the joint WTD does not factorize into a product of the individual WTDs, $\mathcal{W}_{I}(\tau_1,\tau_2)\neq \mathcal{W}_{I}(\tau_1)\mathcal{W}_{I}(\tau_2)$. This shows us that subsequent electron waiting times are correlated if we have no knowledge about the external driving of the capacitor. These classical correlations are due to the regular emission of electrons.

\subsection{Non-ideal operating conditions}
Next, we turn to the electron waiting times for a mesocopic capacitor which is operated under non-ideal conditions. In this case, cycle-missing events may occur where the capacitor fails to emit an electron (or a hole) within a period. The probability that the mesoscopic capacitor (if filled) emits an electron within the first period is given by the normalization factor in Eq.~\eqref{eq:wavmescap}
\begin{equation}
\label{eq:success}
\sigma=1-e^{-\frac{\mathcal{T}}{2\tau_D}}.
\end{equation}
For $\tau_D \ll \mathcal{T}$, the probability approaches unity as one would expect. The probability that the capacitor is refilled within a period is also $\sigma$. To find the ITP, we introduce the probability $P_1$ for the capacitor to be filled at the beginning of a period. This probability fulfills
\begin{equation}\label{eq:rateequations}
\begin{aligned}
P_1(t=0)=&[1-P_1(t=-\mathcal{T})]\sigma+P_1(t=-\mathcal{T})\left(\sigma^2+1-\sigma\right)\\
=&P_1(t=0)\left(\sigma-1\right)^2+\sigma,
\end{aligned}
\end{equation}
having used $P_1(t=0)=P_1(t=-\mathcal{T})$. The first line is the sum of the probabilities that the initially empty capacitor is refilled with probability $\sigma$ or the initially occupied capacitor is emptied and refilled (factor of $\sigma^2$) or is never emptied (factor of $1-\sigma$). We then find
\begin{equation}
\label{eq:p1p0}
P_1(0)=\frac{1}{2-\sigma}=\frac{1}{1+e^{-\frac{\mathcal{T}}{2\tau_D}}}.
\end{equation}
We will also need the probability for the capacitor to be empty at time $t^s$ (again taken within the first period)
\begin{equation}
\label{eq:p0ts}
P_0(t^s)=P_0(0)+P_1(0)\sigma\int\limits_{0}^{t^s}\left|\psi(t)\right|^2dt,
\end{equation}
which includes the probability that the capacitor is emptied during the interval $[0,t^s]$.

In terms of these probabilities, the ITP can be written as
\begin{equation}
\label{eq:itpreal}
\begin{aligned}
\Pi(t^s,t^e)=&e^{-n\frac{\mathcal{T}}{2\tau_D}}[P_0(t^s)+P_1(0)\alpha_n(t^e)]\\&+n\sigma e^{-(n-1)\frac{\mathcal{T}}{2\tau_D}}\alpha_n(t^e)P_0(t^s),
\end{aligned}
\end{equation}
for $t^e\in[n\mathcal{T},(n+1)\mathcal{T}]$, having introduced the probability
\begin{equation}
\alpha_n(t^e)=1-\sigma\int\limits_{n\mathcal{T}}^{t^e}|\psi(t-n\mathcal{T})|^2 dt,
\end{equation}
that the occupied capacitor at $t=n\mathcal{T}$ remains occupied until $t^e$. Only the second line of Eq.~\eqref{eq:itpreal} is a product of terms depending on $t^s$ and $t^e$ which thus will contribute to the WTD. For the first passage time distribution we find using Eq.~\eqref{eq:fptd1}
\begin{equation}
\label{eq:fptdreal}
\begin{aligned}
\mathcal{F}(t^s,t^e)=&\sum\limits_{n=0}^{\infty}\sigma|\psi(t^e-n\mathcal{T})|^2\bigg[e^{-n\frac{\mathcal{T}}{2\tau_D}}P_1(0)\\&+n\sigma e^{-(n-1)\frac{\mathcal{T}}{2\tau_D}}P_0(t^s)\bigg].
\end{aligned}
\end{equation}
Differentiating this expression with respect to $t^s$ [cf.~Eq.~\eqref{eq:wtditp1}] the WTD becomes
\begin{equation}
\label{eq:wtdrealtt}
\begin{aligned}
I(t^s)&\mathcal{W}(t^s,t^e)=P_1(0)\sigma^3\\&\times\sum\limits_{n=0}^{\infty}(n+1)e^{-n\frac{\mathcal{T}}{2\tau_D}}I_{I}(t^s)\mathcal{W}_{I}(t^s,t^e-n\mathcal{T}).
\end{aligned}
\end{equation}
This is a sum of the WTD in Eq.~(\ref{eq:ttwtdid}) for ideal operating conditions, shifted by multiples of the period and rescaled by a factor that accounts for cycle missing events. By integrating over one period of the driving, we find
\begin{equation}
\label{eq:wtdreal}
\mathcal{W}(\tau)=\sigma^2\sum\limits_{n=0}^{\infty}(n+1)e^{-n\frac{\mathcal{T}}{2\tau_D}}\mathcal{W}_{I}(\tau-n\mathcal{T})
\end{equation}
with the mean waiting time
\begin{equation}
\langle\tau\rangle=\frac{\mathcal{T}}{\sigma P_1(0)}=\frac{1+e^{-\frac{\mathcal{T}}{2\tau_D}}}{1-e^{-\frac{\mathcal{T}}{2\tau_D}}}\mathcal{T}.
\end{equation}

Finally, upon inserting Eq.~\eqref{eq:otwtdid} into Eq.~\eqref{eq:wtdreal} and performing the summation, we arrive at the WTD
\begin{equation}
\label{eq:wtdrealexpl}
\mathcal{W}(\tau)=\frac{n(\tau)}{\tau_D}e^{-n(\tau)\frac{\mathcal{T}}{2\tau_D}}\sinh\left(\frac{\mathcal{T}}{2\tau_D}-\frac{\left|\tau-n(\tau)\mathcal{T}\right|}{\tau_D}\right),
\end{equation}
where
\begin{equation}
n(\tau)=\left\lfloor\frac{\tau+\mathcal{T}/2}{\mathcal{T}}\right\rfloor,
\end{equation}
with $\lfloor\cdot\rfloor$ denoting the floor function. This result has previously been derived in Ref.~\cite{albert11} using a rate equation for the occupation probability of the mesoscopic capacitor (correcting here for a misprint~\cite{correction}). It is not surprising that the two approaches yield the same result. The wave packet approach only uses the particle densities, not the wave functions, and we could thus expect that we would recover the result from a classical rate equation description.

\begin{figure}
  \centering
  \includegraphics[width=\columnwidth]{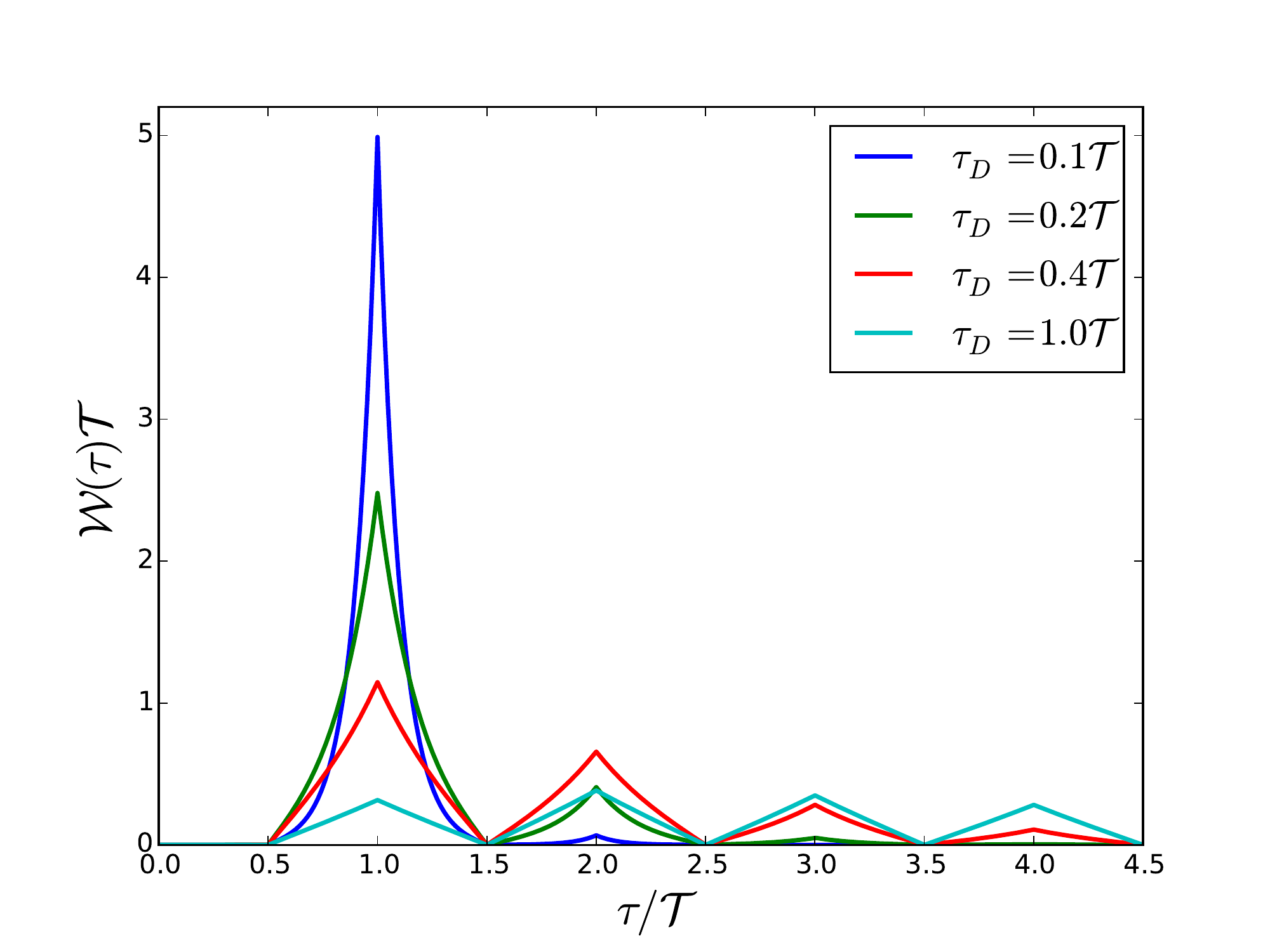}
  \caption{(color online). Analytic results for the WTD of the mesoscopic capacitor. The capacitor is operated under non-ideal conditions, where cycle missing events may occur. We show results for different dwell times $\tau_D$. For a given dwell time, each peak in the WTD has the same shape but is rescaled by a factor that reflects the quality of the capacitor as a single-electron emitter.}
  \label{fig:wtdreal}
\end{figure}

In Fig.~\ref{fig:wtdreal} we show WTDs for the mesoscopic capacitor with different dwell times. For $\tau_D\ll\mathcal{T}$, the capacitor emits one electron and one hole in nearly every cycle and the WTD is strongly peaked around the period of the driving. As the dwell time is increased, the capacitor may fail to emit within a period. These cycle-missing events are clearly visible as peaks in the WTD at multiples of the period. As the dwell time is increased beyond the period of the driving, the synchronization between the external drive and the emission of particles is eventually lost.

\begin{figure}
  \centering
  \includegraphics[width=0.95\columnwidth]{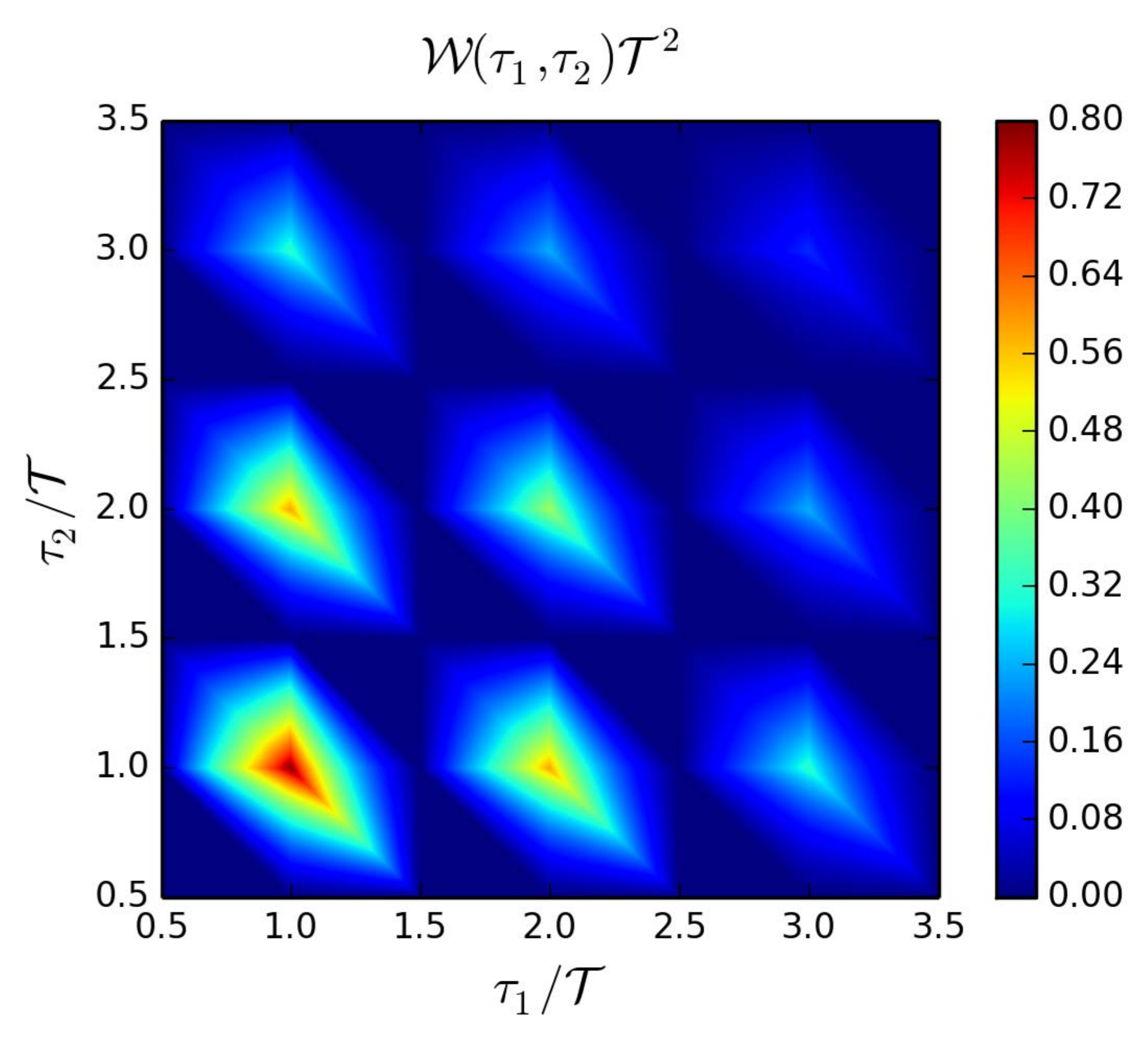}
  \caption{(color online). Analytic result for the joint distribution of waiting times. The capacitor is operated under non-ideal conditions, where cycle missing events may occur. The joint WTD is made up of non-overlapping peaks which are rescaled by a certain factor due to cycle missing event. The dwell time is $\tau_D=0.4\mathcal{T}$.}
  \label{fig:wtd2tnonid}
\end{figure}

Next, we evaluate the joint distribution of waiting times. To this end, we first consider the probability of only having a single detection at time $t^m$ within the interval $[t^s,t^e]$. Ignoring terms which do not depend on both $t^s$ and $t^e$, we find
\begin{equation}
\label{eq:itpfor2twtd}
\begin{aligned}
&-\partial_{{t}_2^e}\left.\Pi(t^s,t^e;{t}_2^s,{t}_2^e)\right|_{{t}_2^s={t}_2^e=t^m}=\left<\hat{b}^\dag(t^m):e^{-\widehat{Q}}:\hat{b}(t^m)\right>\\
&=e^{\frac{\mathcal{T}}{2\tau_D}}\sigma^2\frac{n_1n_2}{n_1+n_2}\left|\psi(t^m-n_1\mathcal{T})\right|^2\Pi(t^s,t^e),
\end{aligned}
\end{equation}
if $t^m\in[n_1\mathcal{T},(n_1+1)\mathcal{T}]$ and $t^e\in[(n_1+n_2)\mathcal{T},(n_1+n_2+1)\mathcal{T}]$. Here $\sigma$ is given by Eq.~\eqref{eq:success} and $\Pi(t^s,t^e)$ by Eq.~\eqref{eq:itpreal}. Differentiating this expression with respect to $t^s$ and $t^e$ according to Eq.~\eqref{eq:wait2t} and integrating over $t^s$ according to Eq.~\eqref{eq:wtd2taveraged} we then find the joint WTD
\begin{equation}
\label{eq:wait2treal}
\begin{aligned}
\mathcal{W}(\tau_1,\tau_2)&=\sigma^4\sum_{n_1,n_2=0}^{\infty}(n_1+1)(n_2+1)\\\times&e^{-(n_1+n_2)\frac{\mathcal{T}}{2\tau_D}}\mathcal{W}_I(\tau_1-n_1\mathcal{T},\tau_2-n_2\mathcal{T}).
\end{aligned}
\end{equation}
As an important check, we recover Eq.~\eqref{eq:wtdreal} by integrating over the second waiting time, $\mathcal{W}(\tau)=\int_0^{\infty}\mathcal{W}(\tau,\tau')d\tau'$. The joint WTD is shown in Fig.~\ref{fig:wtd2tnonid} for non-ideal operating conditions where cycle-missing events may occur.

\section{Floquet scattering theory}
\label{sec:floquet}

Having evaluated the WTD analytically for the mesoscopic capacitor within the wave packet approach, we now proceed with a full Floquet scattering matrix calculation. This approach captures all aspects of the non-interacting many-body problem, including the fermionic statistics and the possible overlap of single-particle wave-functions. We calculate the WTD  using the method developed in Ref.~\cite{dasenbrook14}.

The ITP can be written as the Fredholm determinant
\begin{equation}
  \label{eq:itpdetformula}
  \Pi(t^s,t^e) = \operatorname{det} (1 - Q_{t^s,t^e}),
\end{equation}
where the matrix $Q_{t^s,t^e}$, corresponding to the operator in Eq.~(\ref{eq:qoperator}), has the elements
\begin{align}
  \label{eq:qfloquetmatrixelements}
  [Q_{t^s,t^e}]_{\epsilon,\epsilon'} = &\mathop{\mathop{\sum^\infty
      \sum^\infty}_{m=-\lfloor \epsilon/\hbar \Omega \rfloor}}_{n=-\lfloor \epsilon^{'}/\hbar \Omega
    \rfloor} \mathcal{S}_F^\ast(\epsilon,\epsilon_m) \mathcal{S}_F(\epsilon',\epsilon'_n) \nonumber\\
  &\times K_{t^s,t^e}(\epsilon_m,\epsilon_n') \Theta(-\epsilon) \Theta(-\epsilon'),
\end{align}
and the kernel reads
\begin{equation}
  K_{t^s,t^e}(\epsilon,\epsilon') = \frac{1}{\pi} \frac{\sin[(t^e-t^s) (\epsilon-\epsilon')/2]}{\epsilon-\epsilon'} e^{i(t^s+t^e)(\epsilon-\epsilon')/2}.
  \nonumber
\end{equation}
The scattering matrix is given by Eq.~\eqref{eq:floquetsmatrix} with the coefficients $c_n$ for the potential in Eq.~\eqref{eq:potentialstep} obtained from Eq.~\eqref{eq:dynamicphasefourierseries}.

The numerical calculations are demanding. We discretize the kernel in the energy windows $[-n \hbar \Omega, -(n+1) \hbar \Omega]$ with $n \in \mathbf{N}$ using a five-point Gauss-Legendre quad\-rat\-ure rule following a recently developed method to evaluate Fredholm determinants \cite{bornemann10}. To calculate the matrix $Q_{t^s,t^e}$, we have to sum over all Floquet scattering amplitudes. We find that we can cut off the summation
with about $n_\text{max} \approx 2 \Delta / (\hbar \Omega)$ amplitudes. This is the maximum number of energy quanta that a scattered particle can absorb or emit. The determinant is then taken over all the energies in the $n_\text{max}$ compartments. Finally, we
integrate numerically the ITP over one period of the driving and evaluate the WTD according to Eq.~\eqref{eq:wtditp}.

\begin{figure}
  \centering
  \includegraphics[width=\columnwidth]{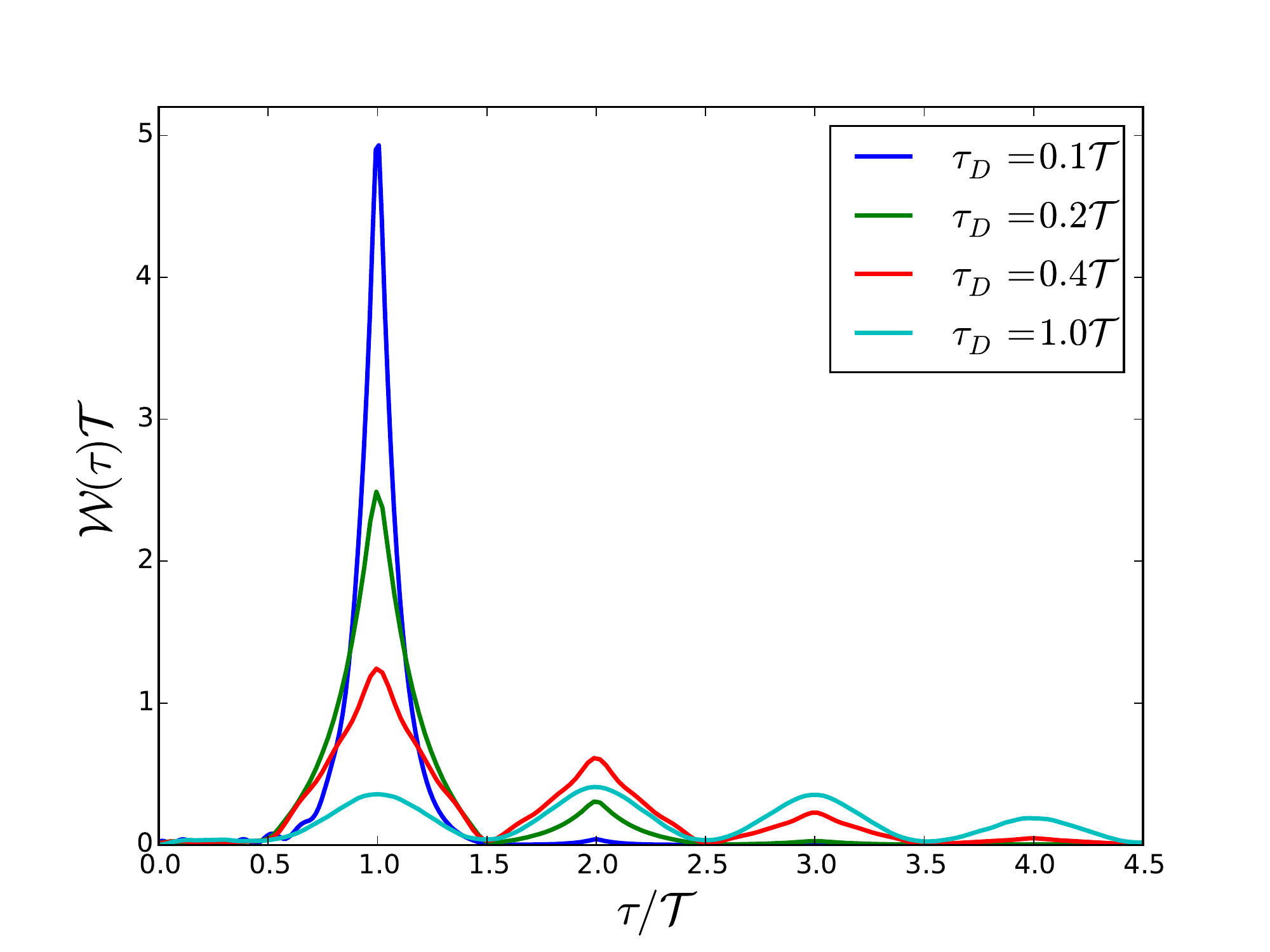}
  \caption{(color online). Floquet calculations of the WTD. Results are shown for different values of the dwell time $\tau_D$. The QPC transmission is $T=0.2$. The results agree very well with those obtained from the wave packet approach in Fig.~\ref{fig:wtdreal}.}
  \label{fig:wtdfloquet}
\end{figure}

Figure~\ref{fig:wtdfloquet} shows WTDs for the mesoscopic capacitor obtained using Floquet scattering theory. The agreement between the Floquet calculations and the wave packet approach is remarkable. For $\tau_D\ll\mathcal{T}$, the central peak at the period of the driving is clearly reproduced with small but visible satellite peaks at multiples of the period. As the dwell time is increased by lowering the level spacing~$\Delta$, the peaks become smoother and less sharp compared to the wave packet approach. This happens as the broadened energy levels start to overlap. This effect is not included in the wave packet approach, where the WTD depends only on the ratio of the dwell time over the period. The dwell time is given by the product of the level spacing and the QPC transmission according to Eq.~(\ref{eq:dwell}). By contrast, in the full scattering problem the level spacing and the transmission are independent parameters which thus provide an additional time scale in the problem. Indeed, a calculation of the finite-frequency noise of the driven capacitor \cite{parmentier12} has shown that the noise vanishes at measurement frequencies that are higher than the level spacing, unlike what is found based on the rate equation description.

\begin{figure}
  \centering
  \includegraphics[width=\columnwidth]{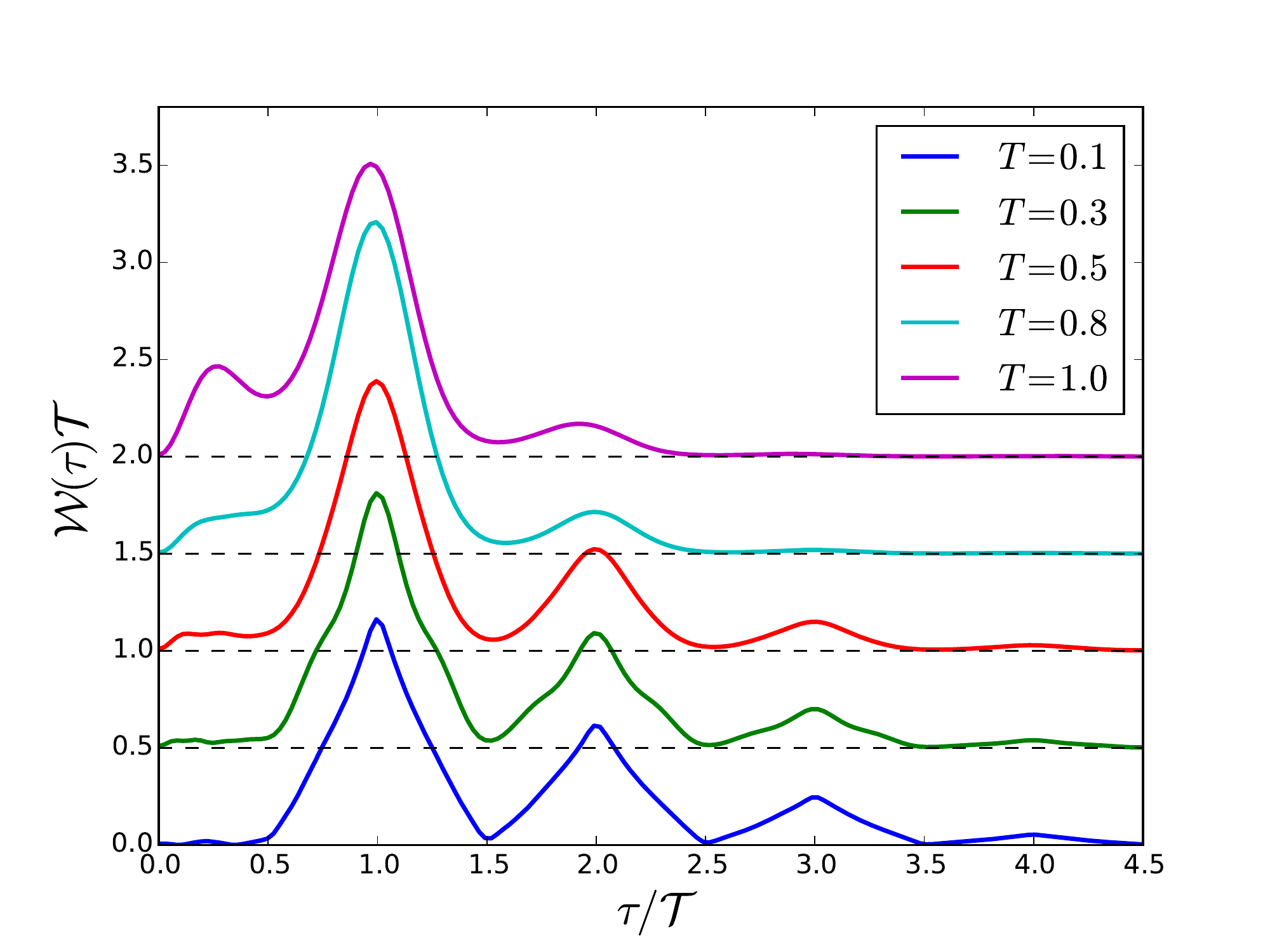}
  \caption{(color online). Floquet calculations of the WTD for different QPC transmission. The dwell time is $\tau_D=0.4 \mathcal{T}$. The different curves have been shifted vertically for the sake of better visibility. These results cannot be captured by the wave packet approach.}
  \label{fig:wtdtransmission}
\end{figure}

With this in mind, we show in Fig.~\ref{fig:wtdtransmission} distributions of waiting times for a fixed dwell time, but with different transmissions of the QPC. As the transmission is increased, the energy levels of the capacitor are broadened and the peaks in the WTD get smeared out. In addition, the peaks at multiples of the period are reduced, as it is increasingly likely that the capacitor will emit an electron in each period. Even with a large transmission, the mesoscopic capacitor seems to function well as a  single-electron emitter. In the extreme case of full transmission, the capacitor consists merely of an elongation of the edge state and the level
quantization is completely lost. It may then happen that more than one electron is lifted above the Fermi level within one period, giving rise to the satellite peak at $\tau\approx\mathcal{T}/4$, similarly to what has been found for Lorentzian-shaped voltage pulses~\cite{dasenbrook14,albert14}.

\section{\label{sec:conc} Conclusions}

We have investigated the distribution of waiting times between the emissions of single electrons from a driven mesoscopic capacitor. Using a wave packet description we have calculated analytically both the electronic waiting time distribution and the joint distribution of subsequent electron waiting times. In the appropriate parameter ranges, these results compare well with full numerical calculations based on Floquet scattering theory. The Floquet method allows us also to calculate waiting time distributions as the transmission between the capacitor and the external reservoirs approaches unity. The agreement between the wave packet approach and the Floquet scattering theory at low transmissions may indicate that the dynamics in this regime essentially is determined by the charge occupation of the capacitor. As such, an indirect measurement of the WTD might be possible by monitoring the charge on the capacitor using for instance an additional quantum point contact.

\section*{Acknowledgments}
We thank M.~Albert, G.~F{\`e}ve, G.~Haack, M.~Moskalets, and P.~Samuelsson for useful discussions. The Floquet calculations were performed using computer resources within the Aalto University School of Science ``Science-IT'' project \cite{aalto-it}. CF is affiliated with Centre for Quantum Engineering at Aalto University. DD gratefully acknowledges the hospitality of Aalto University. PH gratefully acknowledges the hospitality of McGill University. The work was supported by Academy of Finland and Swiss NSF.

\appendix
\section{\label{app:stdist}}
The ITP and first passage time distribution mentioned in Section \ref{sec:ideal} read
\begin{align}
\nonumber
&\Pi_I(\tau)=\left(1-\frac{\tau}{\mathcal{T}}\right)\Theta(\mathcal{T}-\tau)+\frac{\Theta\left(\tau-\frac{\mathcal{T}}{2}\right)\Theta\left(\frac{3\mathcal{T}}{2}-\tau\right)}{\mathcal{T}\left(1-e^{-\frac{\mathcal{T}}{2\tau_D}}\right)^2}\\&\times e^{-\frac{\mathcal{T}}{2\tau_D}}\left[\tau_D\sinh\left(\frac{\frac{\mathcal{T}}{2}-|\tau-\mathcal{T}|}{\tau_D}\right)+|\tau-\mathcal{T}|-\frac{\mathcal{T}}{2}\right]\nonumber
\end{align}
and
\begin{align}
\nonumber
&\mathcal{F}_I(\tau)=\frac{1}{\mathcal{T}}\Theta(\mathcal{T}-\tau)+\frac{\Theta\left(\tau-\frac{\mathcal{T}}{2}\right)\Theta\left(\frac{3\mathcal{T}}{2}-\tau\right)}{\mathcal{T}\left(1-e^{-\frac{\mathcal{T}}{2\tau_D}}\right)^2}\\&\times {\rm sign}(\mathcal{T}-\tau)e^{-\frac{\mathcal{T}}{2\tau_D}}\left[1-\cosh\left(\frac{\frac{\mathcal{T}}{2}-|\tau-\mathcal{T}|}{\tau_D}\right)\right].\nonumber
\end{align}


\section*{References}

\begin{thebibliography}{10}
\expandafter\ifx\csname url\endcsname\relax
  \def\url#1{\texttt{#1}}\fi
\expandafter\ifx\csname urlprefix\endcsname\relax\def\urlprefix{URL }\fi
\expandafter\ifx\csname href\endcsname\relax
  \def\href#1#2{#2} \def\path#1{#1}\fi

\bibitem{buttiker85}
M.~B\"uttiker, Y.~Imry, R.~Landauer, and S.~Pinhas,``Generalized many-channel
  conductance formula with application to small rings", Phys. Rev. B {\bf 31} (1985)
  6207.

\bibitem{buttiker86}
M.~B\"uttiker, ``Four-Terminal Phase-Coherent Conductance", Phys. Rev. Lett. {\bf 57}
  (1986) 1761.

\bibitem{buttiker86b}
M.~B\"uttiker, ``Role of quantum coherence in series resistors", Phys. Rev. B {\bf 33}
  (1986) 3020.

\bibitem{buttiker88b}
M.~B\"uttiker, ``Coherent and sequential tunneling in series barriers", IBM J. Res. Dev. {\bf 32} (1988) 63.

\bibitem{buttiker88}
M.~B\"uttiker, ``Absence of backscattering in the quantum Hall effect in
  multiprobe conductors", Phys. Rev. B {\bf 38} (1988) 9375.

\bibitem{buttiker88c}
M.~B\"uttiker, ``Negative resistance fluctuations at resistance minima in narrow
  quantum Hall conductors", Phys. Rev. B {\bf 38} (1988) 12724.

\bibitem{buttiker93}
M.~B\"uttiker, A.~Pr\^etre, and H.~Thomas, ``Dynamic Conductance and the Scattering
  Matrix of Small Conductors", Phys. Rev. Lett. {\bf 70} (1993) 4114.

\bibitem{moskalets02}
M.~Moskalets and M.~B\"uttiker, ``Floquet scattering theory of quantum pumps", Phys.
  Rev. B {\bf 66} (2002) 205320.

\bibitem{buttiker90}
M.~B\"uttiker, ``Scattering Theory of Thermal and Excess Noise in Open
  Conductors", Phys. Rev. Lett. {\bf 65} (1990) 2901.

\bibitem{buttiker92}
M.~B\"uttiker, ``Scattering theory of current and intensity noise correlations in
  conductors and wave guides", Phys. Rev. B {\bf 46} (1992) 12485.

\bibitem{blanter00}
Ya.~M. Blanter and M.~B\"uttiker, ``Shot noise in mesoscopic conductors", Phys. Rep.
  {\bf 336} (2000) 1.

\bibitem{pilgram03}
S.~Pilgram, A.~N. Jordan, E.~V. Sukhorukov, and M.~B\"uttiker, ``Stochastic Path
  Integral Formulation of Full Counting Statistics", Phys. Rev. Lett. {\bf 90} (2003)
  206801.

\bibitem{nagaev04}
K.~E. Nagaev, S.~Pilgram, and M.~B\"uttiker, ``Frequency Scales for Current
  Statistics of Mesoscopic Conductors", Phys. Rev. Lett. {\bf 92} (2004) 176804.

\bibitem{samuelsson03}
P.~Samuelsson, E.~V. Sukhorukov, and M.~B\"uttiker, ``Orbital Entanglement and
  Violation of Bell Inequalities in Mesoscopic Conductors", Phys. Rev. Lett. {\bf 91}
  (2003) 157002.

\bibitem{samuelsson04}
P.~Samuelsson, E.~V. Sukhorukov, and M.~B\"uttiker, ``Two-Particle Aharonov-Bohm
  Effect and Entanglement in the Electronic Hanbury Brown--Twiss Setup",
  Phys. Rev. Lett. {\bf 92} (2004) 026805.

\bibitem{foerster08}
H.~F\"orster and M.~B\"uttiker, ``Fluctuation Relations without Microreversibility
  in Nonlinear Transport", Phys. Rev. Lett. {\bf 101} (2008) 136805.

\bibitem{sanchez10}
R.~S\'anchez, R.~L\'opez, D.~S\'anchez, and M.~B\"uttiker, ``Mesoscopic Coulomb Drag,
  Broken Detailed Balance, and Fluctuation Relations", Phys. Rev. Lett. {\bf 104}
  (2010) 076801.

\bibitem{sanchez11}
R.~S\'anchez and M.~B\"uttiker, ``Optimal energy quanta to current conversion", Phys.
  Rev. B {\bf 83} (2011) 085428.

\bibitem{bergenfeldt14}
C.~Bergenfeldt, P.~Samuelsson, B.~Sothmann, C.~Flindt, and M.~B\"uttiker, ``Hybrid
  Microwave-Cavity Heat Engine", Phys. Rev. Lett. {\bf 112} (2014) 076803.

\bibitem{li12}
J.~Li, G.~Fleury, and M.~B\"uttiker, ``Scattering theory of chiral Majorana fermion
  interferometry", Phys. Rev. B {\bf 85} (2012) 125440.

\bibitem{jacquod13}
P.~Jacquod and M.~B\"uttiker, ``Signatures of Majorana fermions in hybrid
  normal-superconducting rings", Phys. Rev. B {\bf 88} (2013) 241409.

\bibitem{buttiker82}
M.~B\"uttiker and R.~Landauer, ``Traversal Time for Tunneling", Phys. Rev. Lett. {\bf 49}
  (1982) 1739.

\bibitem{buttiker83}
M.~B\"uttiker, ``Larmor precession and the traversal time for tunneling", Phys.
  Rev. B {\bf 27} (1983) 6178.

\bibitem{albert11}
M.~Albert, C.~Flindt, and M.~B\"uttiker, ``Distributions of Waiting Times of Dynamic
  Single-Electron Emitters", Phys. Rev. Lett. {\bf 107} (2011) 086805.

\bibitem{albert12}
M.~Albert, G.~Haack, C.~Flindt, and M.~B\"uttiker, ``Electron Waiting Times in
  Mesoscopic Conductors", Phys. Rev. Lett. {\bf 108} (2012) 186806.

\bibitem{dasenbrook14}
D.~Dasenbrook, C.~Flindt, and M.~B\"uttiker, ``Floquet Theory of Electron Waiting
  Times in Quantum-Coherent Conductors", Phys. Rev. Lett. {\bf 112} (2014) 146801.

\bibitem{buttiker93a}
M.~B{\"u}ttiker, H.~Thomas, and A.~Pr{\^e}tre, ``Mesoscopic capacitors", Phys. Lett. A
  {\bf 180} (1993) 364.

\bibitem{pretre96}
A.~Pr\^etre, H.~Thomas, and M.~B\"uttiker, ``Dynamic admittance of mesoscopic
  conductors: Discrete-potential model", Phys. Rev. B {\bf 54} (1996) 8130.

\bibitem{gopar96}
V.~A. Gopar, P.~A. Mello, and M.~B\"uttiker, ``Mesoscopic Capacitors: A Statistical
  Analysis", Phys. Rev. Lett. {\bf 77} (1996) 3005.

\bibitem{brouwer97}
P.~W. Brouwer and M.~B\"uttiker, ``Charge-relaxation and dwell time in the
  fluctuating admittance of a chaotic cavity", Europhys. Lett. {\bf 37} (1997) 441.

\bibitem{gabelli06}
J.~Gabelli, G.~F{\`e}ve, J.-M. Berroir, B.~Pla{\c{c}}ais, A.~Cavanna,
  B.~Etienne, Y.~Jin, and D.~Glattli, ``Violation of Kirchhoff's laws for a coherent
  {RC} circuit", Science {\bf 313} (2006) 499.

\bibitem{glattli14talk}
D.~C. Glattli, ``Markus {B}\"uttiker memorial talk: From shot noise and ac
  transport to electron quantum optics", invited talk at the 27th International
  Conference on Low Temperature Physics, Buenos Aires, August, 2014.

\bibitem{mora:2010}
C.~Mora and K.~L. Hur, ``Universal resistances of the quantum {RC} circuit", Nat.
  Phys. {\bf 6} (2010) 697.

\bibitem{moskalets08}
M.~Moskalets, P.~Samuelsson, and M.~B\"uttiker, ``Quantized Dynamics of a Coherent
  Capacitor", Phys. Rev. Lett. {\bf 100} (2008) 086601.

\bibitem{parmentier12}
F.~D. Parmentier, E.~Bocquillon, J.-M. Berroir, D.~C. Glattli,
  B.~Pla\ifmmode~\mbox{\c{c}}\else \c{c}\fi{}ais, G.~F\`eve, M.~Albert,
  C.~Flindt, and M.~B\"uttiker, ``Current noise spectrum of a single-particle
  emitter: Theory and experiment", Phys. Rev. B {\bf 85} (2012) 165438.

\bibitem{feve07}
G.~F{\`e}ve, A.~Mahe, J.-M. Berroir, T.~Kontos, B.~Placais, D.~Glattli,
  A.~Cavanna, B.~Etienne, and Y.~Jin, ``An on-demand coherent single-electron source",
  Science {\bf 316} (2007) 1169.

\bibitem{bocquillon13}
E.~Bocquillon, V.~Freulon, J.-M. Berroir, P.~Degiovanni, B.~Pla{\c{c}}ais,
  A.~Cavanna, Y.~Jin, and G.~Feve, ``Coherence and indistinguishability of single
  electrons emitted by independent sources", Science {\bf 339} (2013) 1054.

\bibitem{bocquillon14}
E.~Bocquillon, V.~Freulon, F.~D. Parmentier, J.-M. Berroir, B.~Pla{\c{c}}ais,
  C.~Wahl, J.~Rech, T.~Jonckheere, T.~Martin, C.~Grenier,
  D.~Ferraro, P.~Degiovanni, and G.~F\`eve, ``Electron
  quantum optics in ballistic chiral conductors", Ann. Phys. (Berlin) {\bf 526} (2014) 1.

\bibitem{bocquillon12}
E.~Bocquillon, F.~D. Parmentier, C.~Grenier, J.-M. Berroir, P.~Degiovanni,
  D.~C. Glattli, B.~Pla\ifmmode~\mbox{\c{c}}\else \c{c}\fi{}ais, A.~Cavanna,
  Y.~Jin, and G.~F\`eve, ``Electron Quantum Optics: Partitioning Electrons One by
  One", Phys. Rev. Lett. {\bf 108} (2012) 196803.

\bibitem{brandes08}
T.~Brandes, ``Waiting times and noise in single particle transport", Ann. Phys.
  (Berlin) {\bf 17} (2008) 477.

\bibitem{welack09}
S.~Welack, S.~Mukamel, and Y.~Yan, ``Waiting time distributions of electron transfers
  through quantum dot Aharonov-Bohm interferometers", Europhys. Lett. {\bf 85} (2009)
  57008.

\bibitem{rajabi13}
L.~Rajabi, C.~P\"oltl, and M.~Governale, ``Waiting Time Distributions for the
  Transport through a Quantum-Dot Tunnel Coupled to One Normal and One
  Superconducting Lead", Phys. Rev. Lett. {\bf 111} (2013) 067002.

\bibitem{thomas13}
K.~H. Thomas and C.~Flindt, ``Electron waiting times in non-Markovian quantum
  transport", Phys. Rev. B {\bf 87} (2013) 121405.

\bibitem{albert14}
M.~Albert and P.~Devillard, ``Waiting time distribution for trains of quantized
  electron pulses", Phys. Rev. B {\bf 90} (2014) 035431.

\bibitem{thomas14}
K.~H. Thomas and C.~Flindt, ``Waiting time distributions of noninteracting fermions
  on a tight-binding chain", Phys. Rev. B {\bf 89} (2014) 245420.

\bibitem{haack14}
G.~Haack, M.~Albert, and C.~Flindt, ``Distributions of electron waiting times in
  quantum-coherent conductors", Phys. Rev. B {\bf 90} (2014) 205429.

\bibitem{sothman14}
B.~Sothmann, ``Electronic waiting-time distribution of a quantum-dot spin valve",
  Phys. Rev. B {\bf 90} (2014) 155315.

\bibitem{tang14}
G.-M. Tang, F.~Xu, and J.~Wang, ``Waiting time distribution of quantum electronic
  transport in the transient regime", Phys. Rev. B {\bf 89} (2014) 205310.

\bibitem{dasenbrook15}
D.~Dasenbrook, P.~P. Hofer, and C.~Flindt, ``Electron waiting times in coherent
  conductors are correlated", Phys. Rev. B {\bf 91} (2015) 195420.

\bibitem{moskalets11book}
M.~Moskalets, ``Scattering Matrix Approach to Non-Stationary Quantum Transport",
  Imperial College Press, 2011.

\bibitem{keeling08}
J.~Keeling, A.~V. Shytov, and L.~S. Levitov, ``Coherent Particle Transfer in an
  On-Demand Single-Electron Source", Phys. Rev. Lett. {\bf 101} (2008) 196404.

\bibitem{moskalets13}
M.~Moskalets, G.~Haack, and M.~B\"uttiker, ``Single-electron source: Adiabatic versus
  nonadiabatic emission", Phys. Rev. B {\bf 87} (2013) 125429.

\bibitem{mahe10}
A.~Mah\'e, F.~D. Parmentier, E.~Bocquillon, J.-M. Berroir, D.~C. Glattli,
  T.~Kontos, B.~Pla\ifmmode~\mbox{\c{c}}\else \c{c}\fi{}ais, G.~F\`eve,
  A.~Cavanna, and Y.~Jin, ``Current correlations of an on-demand single-electron
  emitter", Phys. Rev. B {\bf 82} (2010) 201309.

\bibitem{albert10}
M.~Albert, C.~Flindt, and M.~B\"uttiker, ``Accuracy of the quantum capacitor as a
  single-electron source", Phys. Rev. B {\bf 82} (2010) 041407.

\bibitem{moskalets13noise}
M.~Moskalets, ``Noise of a single-electron emitter", Phys. Rev. B {\bf 88} (2013)
  035433.

\bibitem{haake}
F. Haake, ``Quantum Signatures of Chaos", Springer-Verlag, 2001.

\bibitem{sasaoka10}
K.~Sasaoka, T.~Yamamoto, and S.~Watanabe, ´´Single-electron pumping from a quantum
  dot into an electrode", Appl. Phys. Lett. {\bf 96} (2010) 102105.

\bibitem{correction}
Below Eq.~(6) of Ref.~\cite{albert11}, the quantity $\xi_{ee}$ should be defined as $\xi_{ee}=\Gamma [\Delta t-\lfloor\Delta t/T+1/2\rfloor T]$.

\bibitem{bornemann10}
F.~Bornemann, ``On the numerical evaluation of Fredholm determinants", Math. Comp.
  {\bf 79} (2010) 871.

\bibitem{aalto-it}
\url{http://science-it.aalto.fi/}.

\end{thebibliography}
\end{document}